\documentclass{optica-article}
\journal{opticajournal}
\articletype{Research Article}
\usepackage{hyperref}
\usepackage{lineno}
\usepackage[separate-uncertainty,detect-all]{siunitx}

\DeclareSIUnit{\Molar}{M}

\newcommand{\uM}{\micro\Molar}
\newcommand{\mM}{\milli\Molar}
\newcommand{\mgdl}{\mg\per\deci\liter}

\graphicspath{{figures/}}

\begin{document}

\title{A Quantitative Holographic Agglutination Assay for Immunoglobulin A}

\author{Rushna Quddus,\authormark{1,2,*} Kent Kirshenbaum,\authormark{1} and David G. Grier\authormark{3}}

\address{\authormark{1}Department of Chemistry, New York University, New York, NY 10003, USA\\
\authormark{2}Currently with Adolphe Merkle Institute, University of Fribourg, CH-1700 Fribourg, Switzerland \\
\authormark{3}Department of Physics and Center for Soft Matter Research, New York University, New York, NY 10003, USA}

\email{\authormark{*}rushna.quddus@unifr.ch} 

\begin{abstract*}
This study introduces a Holographic Agglutination Assay for quantifying levels of the immunoglobulin protein IgA in biological samples.
This is the first example of a label-free and bead-free assay that 
quantifies protein agglutinates by direct detection
using Total Holographic Characterization.
A proof-of-concept assay for human serum immunoglobulins is demonstrated using Jacalin, the galactose-specific plant lectin, to induce selective agglutination. By analyzing the size, refractive index and number of particles in an assay sample, we obtain a reproducible and quantitative measurement of galactosylated immunoglobulins in a given sample.
The assay is calibrated for a physiologically relevant reference interval of IgA concentrations in a 10\texttimes-diluted emulated biological sample from low (\qty{80}{\mgdl}, \qty{5}{\uM}) to high (\qty{320}{\mgdl}, \qty{20}{\uM}) levels.
The assay clearly distinguishes samples containing IgA from samples containing IgG.
More broadly, this study introduces a
platform for creating lectin-mediated Holographic Agglutination Assays to monitor levels of immunoglobulins in biological samples.
The ability to quantify immunoglobulin levels efficiently in clinical samples 
is likely to be valuable for diagnostics
and will provide a basis for assaying other proteins that can be induced to agglutinate.
\end{abstract*}

\section{Introduction}

\begin{figure}[ht]
\centering
  \includegraphics[width=\textwidth]{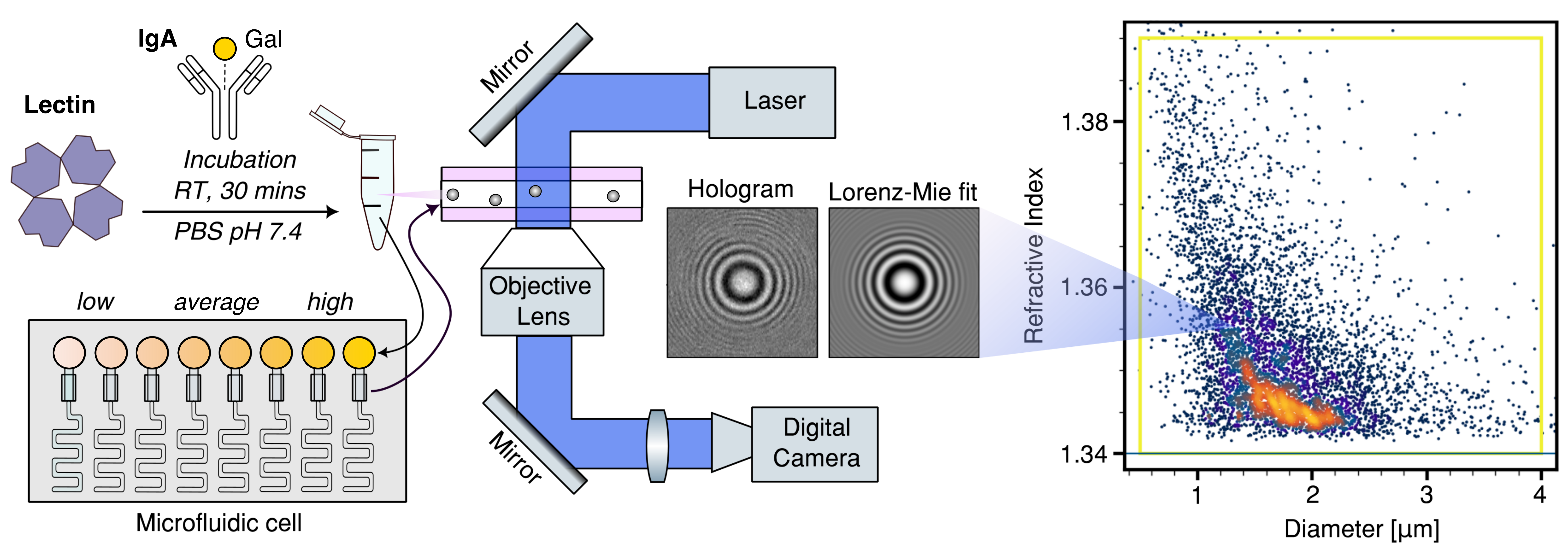}
  \caption{Schematic of the Holographic Agglutination Assay.
  The agglutination assay is initiated by incubating the tetrameric protein, lectin, with the glycoprotein immunoglobulin A (upper left) in a microfuge tube. Gal represents the O-linked galactose on IgA that binds to the lectin Jacalin to form protein agglutinates. The microfluidic cell shown (lower left) can be used to quantify the Human serum IgA levels in each sample well. The optical setup (middle) used for Total Holographic Characterization to produce and record holograms of the agglutinates as they flow through the microfluidic channel.  Representative distribution plot (right) of all detected particles in a single measurement depicting the values of diameter and refractive index from the fitted holograms.}
  \label{fig:schematic}
\end{figure}

A hologram of a micrometer-scale object 
encodes a wealth of information about the object's size, 
shape and composition as well as its three-dimensional position.
This information can be extracted with great precision
by fitting a recorded hologram to a
suitable generative model for the image-formation
process \cite{lee2007characterizing,martin2022inline}. 
This technique is called Total Holographic Characterization (THC).
THC can determine the diameter of a colloidal particle with nanometer precision and can resolve its refractive index at the imaging wavelength to within a part per thousand \cite{krishnatreya2014measuring}.
Although this technique was originally developed for use with homogeneous colloidal spheres, it can also be applied
to inhomogeneous \cite{cheong2011holographic,odete2020role,wang2016holographic,altman2020interpreting} and aspherical
\cite{fung2013holographic,wang2014using,wang2016holographic,wang2016protein,altman2021holographic,altman2023anomalous}
particles, either by adopting a
generative model that
accounts for such particles' complex light-scattering properties \cite{fung2013holographic,wang2014using}
or by interpreting the results of the
standard implementation 
with effective-medium theory
\cite{cheong2011holographic,altman2020interpreting,odete2020role}.
When applied to micrometer-scale protein aggregates (as distinct from protein agglutinates),
the effective-sphere interpretation of THC provides reliable estimates
for the aggregates' size distribution
\cite{winters2020quantitative},
concentration \cite{wang2016protein},
and fractal dimension \cite{wang2016holographic,abdulali2022multi}. 
In addition, THC can usefully distinguish
protein aggregates from other co-dispersed particles on the basis
of their refractive index
\cite{winters2020quantitative}. 

THC has previously been used to implement label-free bead-based molecular binding assays for antibodies in solution
\cite{zagzag2020holographic,snyder_quddus_2020}.
These tests first immobilize the receptor protein
on the surface of suitably functionalised
colloidal spheres.
THC then detects
the presence of selectively bound target proteins by
measuring nanometre-scale changes in the beads' diameters.
Bead-based holographic binding assays work with
microliter sample volumes and report the concentration of the target macromolecule in a matter of minutes.
In this report, we introduce 
bead-free holographic assays 
that work by inducing target proteins to
form micrometre-scale agglutinates that are
detected and characterized with THC. 
Relative to bead-based assays, holographic agglutination
assays eliminate
the time, effort and cost of developing and using
probe beads and mitigates uncertainties associated
with nonspecific binding to the substrate beads.

\subsection{Total Holographic Characterization}

Total Holographic Characterisation (THC) uses in-line holographic video microscopy to measure properties of individual colloidal particles
in a colloidal dispersion \cite{lee2007characterizing}.
As depicted schematically in Fig.~\ref{fig:schematic}, the instrument transports the sample
down a microfluidic channel, where it is illuminated by a collimated laser beam.
Light scattered by a particle interferes with the illuminating beam to form
a hologram of the particle whose magnified intensity distribution is recorded with a video camera.
Each recorded hologram is fit to a generative model based on the Lorenz-Mie theory
of light scattering to estimate the particle's apparent diameter, $d_p$, and refractive index, $n_p$.
THC is the only particle characterization technique
that can measure individual particles' refractive indexes
\emph{in situ}\cite{cheong2009flow}.
This capability is useful for differentiating
populations of particles by composition and has
been the inspiration for a host of applications involving
materials such as oil droplets\cite{kasimbeg_holographic_2019},
colloidal spheres\cite{middleton_optimizing_2019},
protein aggregates\cite{rahn2023strengths} and microorganisms\cite{bedrossian_quantifying_2017}.

The scatter plot in Fig.~\ref{fig:schematic}
presents THC results for a typical sample,
with each point representing the
diameter, $d_p$, and refractive index, $n_p$, of
one particle.
The points are colored by the local density of observations from high (orange) to low (blue).
Because THC records a hologram of each particle in a fixed volume
of fluid, it 
accurately measures the concentration
of dispersed particles in the total sample.
This approach works for concentrations ranging
from \qty{e3}{particles\per\milli\liter}
to \qty{e8}{particles\per\milli\liter}.
The lower limit is set by counting statistics; 
concentrations greater than \qty{e8}{particles\per\milli\liter} cause excessive
overlapping of the holograms of neighboring particles.

\begin{figure}[ht]
    \centering
  \includegraphics[width=\textwidth]{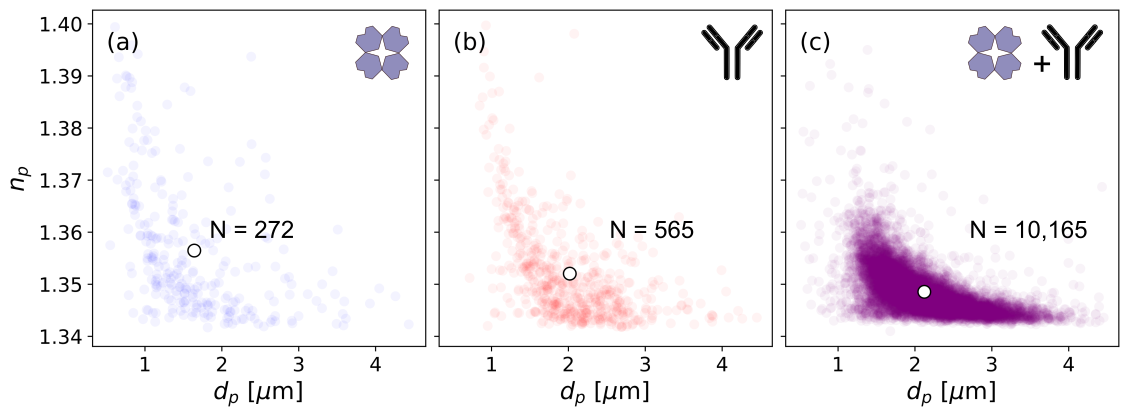}
  \caption{Proof-of-concept demonstration of an agglutination assay for human IgA. Scatter plots show the diameter ($d_p$) and refractive index ($n_p$) of each particle in a sample. 
  (a) Background: Jacalin (\qty{1}{\uM} in PBS buffer, \qty{2}{\micro\liter} sample).
  (b) Background: human IgA (\qty{0.5}{\uM} in PBS buffer, \qty{2}{\micro\liter} sample).
  (c) Assay: Jacalin (\qty{1}{\uM}) incubated with human IgA (\qty{0.2}{\uM}) for \qty{30}{\minute} in PBS buffer, \qty{0.75}{\micro\liter} sample). Each data point represents a single detected particle. Particle count increases more than 10-fold due to agglutination of lectin and IgA.  White circles denote mean values of the distributions.}
	\label{fig:scatter}
\end{figure}

Previous studies have used THC for direct
read-out of bead-based immunoassays \cite{luo_quantitative_2021,polpanich2007detection}.
One approach uses functionalized probe beads to bind the
target analyte and then relies on THC to monitor the associated
change in the beads' diameter\cite{cheong2009flow,zagzag2020holographic,snyder_quddus_2020}.
An alternative approach relies on analyte molecules to mediate aggregation of functionalized probe beads
\cite{luo_quantitative_2021} and uses machine-learning techniques to estimate analyte concentration
from recorded holograms of the aggregated beads \cite{polpanich2007detection}.
Our approach dispenses with probe beads altogether
by directly detecting agglutination of proteins using the measured
distribution of agglutinate size and refractive index
to estimate the total concentration of the target protein ligands.

\subsection{Agglutination Assays}
Protein agglutinates form in a process that typically is mediated by specific interactions
such as those between antibodies and antigens \cite{stavitsky_agglutination_1998}. Unlike protein aggregates produced by non-specific intermolecular association, 
protein agglutinates are formed by specific multivalent binding interactions between proteins 
to produce visible dispersed particles\cite{arrhenius_agglutination_1908}.
Agglutination of immunoglobulins is a routine test for blood typing \cite{malomgre_recent_2009,li_blood_2022} 
and diagnosing bacterial \cite{widal1896serodiagnostic, olopoenia_widal_2000} or viral infections \cite{hirst1942quantitative, esmail_rapid_2021}. 
Currently, there are no analytical methods known to directly infer the glycome of an organism from its genome. 
Thus, agglutination assays based on glycan-specific binding have played a pivotal role in elucidating various aspects of glycobiology and the biochemical roles of glycans \cite{ribeiro_dot_2013}.
Proteins such as lectins are able to selectively bind to O- or N-linked glycans and are often used in the field of glycobiology to characterise and quantify these interactions \cite{rademacher1988glycobiology}.

Among the most comprehensively studied lectins is Jacalin, which has a particular affinity for O-linked glycoproteins with terminal galactose residues \cite{kabir_Jacalin_1998, arockia_jeyaprakash_structural_2003}. The lectin can be easily obtained from the jackfruit \emph{Artocarpus integrifolia}. 
Jacalin is commonly used to isolate 
galactosylated proteins such as human IgA \cite{roque-barreira_Jacalin_1985}.
When immobilized on a solid-phase support such as agarose resin beads, Jacalin provides the basis for 
affinity purification of IgA \cite{backer_autoantibodies_1989,pack_purification_2000}. 

IgA is the second most abundant antibody isotype in human serum and its medical interest primarily stems from its involvement in nephropathy \cite{andre_impairment_1990} and viral infection \cite{ma_serum_2020}.
Monitoring the concentration of IgA in serum can be critical for early detection and diagnosis 
of disease states and conditions such as immunodeficiencies \cite{allez_immunodeficiency_2004}.
IgA has two subclasses, IgA1 and IgA2 and contains unique O-glycans in its hinge region that terminate in a galactose \cite{baenziger_structure_1974}.
Jacalin is known to preferentially bind to IgA1 \cite{skea_studies_1988}, which is the dominant fraction of human serum\cite{delacroix_iga_1982}.
Notably, human serum immunoglobulin G (IgG) does not generally contain an O-linked terminal galactose 
\cite{parkkinen1989aberrant, alter_antibody_2018, yoann_2016} and so does not bind to Jacalin. 
IgG is therefore useful for demonstrating the selectivity of 
Jacalin agglutination for an immunoglobulin isotype and serves
as a negative control for an IgA agglutination assay.
While several methods are known for detecting glycoproteins and immunoglobulins, 
there is a need for a rapid and quantitative agglutination assay that is not only label-free and bead-free 
but also is reagent-efficient \cite{stavitsky_agglutination_1998}. 
We report a proof-of-concept assay that uses freely dissolved Jacalin to selectively induce agglutination 
of human immunoglobulin A (IgA) from solution. 
This process creates sub-visible colloidal particles that can be directly detected and quantified using THC \cite{lee2007characterizing}.

\section{Materials and Methods}

The commercial instrument for THC used in this study (xSight, Spheryx, Inc.) uses laser light at a vacuum wavelength of \qty{450}{\nm} 
and can characterize particles ranging in
size from \qty{500}{\nm} to \qty{10}{\um}.
A measurement is performed by transferring
up to \qty{30}{\micro\liter} of the sample
into a reservoir in 
a compatible microfluidic chip (xCell8, Spheryx, Inc.).
xSight uses a pressure-driven flow to automatically draw a specified volume
between \qty{0.5}{\micro\liter} and
\qty{6}{\micro\liter} through its observation
volume for analysis.

\subsection{Protein solutions}

All protein solutions are prepared in phosphate-buffered saline (PBS)
at pH~\num{7.4}
with \qty{0.02}{\percent} sodium azide to suppress
bacterial growth and \qty{0.05}{\percent} Tween-20
to prevent solutes from adhering to
the container walls.
Jacalin test solutions are prepared with the commercial
extract from the seeds of \textit{Artocarpus integrifolia} 
(G-Biosciences, lyophilized solid, \qty{10}{\mg}, supplier no.~786-473, VWR catalog no.~71003-186). 
The protein ligands tested for agglutination with Jacalin are human serum IgA
(Sigma-Aldrich, catalog no.~401098-2MG, delivered in
\qty{100}{\mM} NaCl, \qty{100}{\mM} Tris-HCl, pH~8.0).
Human IgG (Sigma-Aldrich, catalog no.~I4506, reagent grade, \qty{95}{\percent}, essentially salt-free, lyophilized powder) was used as a control to compare Jacalin selectivity. 
The assay buffer is composed of 
\qty{10}{\mM} PBS pH 7.4,  
\qty{150}{\mM} NaCl, 
\qty{0.05}{\percent} Tween-20 and \qty{0.02}{\percent} sodium azide.
The storage buffer for Jacalin contains 
\qty{0.1}{\Molar} galactose to preserve the galactose binding sites on the lectin in their native conformational state. The final concentration of galactose in the assay is \qty{10}{\milli\Molar}, which is well below the concentration needed to elute IgA from Jacalin affinity purification columns (\qty{0.5}{\Molar}) or inhibit its agglutination\cite{aucouturier_characterization_1987,arockia_jeyaprakash_structural_2003}. 

\subsection{Holographic Agglutination Assay for IgA}

The experimental protocol is summarized in Fig.~\ref{fig:schematic} and begins by
adding \qty{5}{\micro\liter}, 
of Jacalin solution
to \qty{45}{\micro\liter} of
the protein ligand solution in an microfuge tube. 
The dissolved lectin induces agglutination
in immunoglobulins with the corresponding glycans. 
The concentration of the ligand reported
is the final concentration in \qty{50}{\micro\liter} of the total assay volume after addition of the lectin,
and therefore accounts for dilution by the added test solution. 
The sample is then incubated
at room temperature for \qty{30}{\minute} before being introduced
into one of the sample chip's reservoirs
for automated
analysis by xSight.
Optionally, the incubation may be conducted in the sample reservoir of the microfluidic channel, thereby facilitating time-resolved measurements.  

Agglutinates that grow larger
than \qty{500}{\nm} are
detected and characterized by xSight.
Provided the total concentration of particles falls within the instrument's
operational range, this comprehensive approach to direct counting
yields highly accurate values for the concentration of agglutinates in the
accessible size range.
The holographic agglutination assay is 
designed to produce detectable agglutinates
in the clinically relevant reference interval \cite{whyte2018normal}
of analyte concentrations by optimizing the amount of lectin used. 
For the specific case of serum IgA1, referred to as IgA in the following text, 
concentrations in human serum below \qty{8}{\mgdl} are associated
with IgA deficiency \cite{dieguez2020serum}.
The reference interval  
for adult humans is 
\qty{80}{\mgdl} (physiologically low)
to \qty{320}{\mgdl} (physiologically high).
This corresponds to a concentration range of
\qty{5}{\uM} to \qty{20}{\uM}.
The number of detectable agglutinates in
an agglutination assay for serum IgA
is best matched to the operating range of xSight 
by 10\texttimes\ dilution.
The operating range of the xSight instrument can then cover the
reference interval
with IgA test solutions at
concentrations ranging from
\qty{0.5}{\uM} to \qty{2}{\uM}.
In clinical samples, ten-fold dilution
would also be useful for reducing the background
concentration of pre-existing
dispersed particles.

\begin{figure}
\centering 
\includegraphics[width=0.75\textwidth]{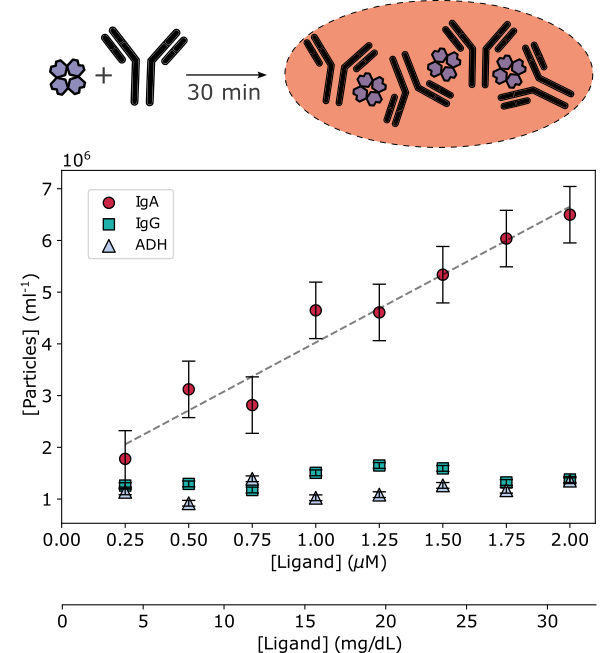}
\caption{Quantitative holographic agglutination assay for
human serum IgA in the presence of Jacalin. Each point in the scatter plot represents the
number density of detected particles as a function of 
ligand concentration for a fixed concentration of the lectin Jacalin,\qty{0.5}{\uM}.
Human IgA (circles) forms agglutinates in the presence of Jacalin
and the number density of agglutinates increases
linearly with IgA concentration over the clinically relevant range.
Human IgG (squares) does not increase the number density of
detected particles above background due to the lack of terminal galactose. 
Similarly, alcohol dehydrogenase (triangles) does not generate a signal above background 
and also serves as a negative control.}
\label{fig:particleconcentration}
\end{figure}

\subsection{Negative control}

Negative control measurements are performed
using the same protocol as for immunoglobulin
agglutination assays.
Alcohol dehydrogenase (ADH1) from yeast (Sigma-Aldrich, catalog no.~A3263-7.5KU) 
is used as a negative control for immunoglobulins based on previous work \cite{snyder_quddus_2020}.
ADH has a comparable molecular weight to a monomeric immunoglobulin (\qty{150}{\kilo\dalton}) but does not contain galactose.
It therefore serves as a negative control for non-specific
agglutination by Jacalin.

\section{Results}
\label{sec:results}

\subsection{Validation of Holographic Agglutination Analysis }

Total Holographic Characterization detects all particles ranging
in size from \qty{500}{\nm} to \qty{10}{\um} that pass through
the observation volume, which extends to the full height
of the microfluidic channel.
Each reported measurement examines a calibrated \qty{3}{\micro\liter} sample volume.
Holographic characterization therefore provides 
more accurate results for particle concentration than
methods such as microflow imaging (MFI), for which the
effective observation volume depends on particle size
and therefore is less well defined \cite{rahn2023strengths}.
Uncertainties in particle concentration are
specified to be smaller than \qty{10}{\percent}
over the range
\qtyrange{e5}{e8}{particles\per\milli\liter}.

A single run through the microfluidic channel at the fixed volume of 3 {$\mu$}L 
provided a total particle count for each sample assayed.
Discrete points 
represent the mean concentration of the distribution of all detected particles 
for each set of conditions.
The reproducibility of the measurement protocol was evaluated by performing biological triplicates 
for each protein measured in different microfluidic cells. 
Error bars report the range of mean particle concentrations obtained
in the triplicate measurements on each of the samples. 

The first demonstration of a holographic agglutination assay for IgA detection is reported in Figure~\ref{fig:scatter}, which presents the raw
data from a typical assay.
The Jacalin and IgA solutions independently
contain a small concentration of detectable
protein aggregates, even before they are mixed and
incubated.
THC analysis on \qty{2}{\micro\liter} sample of 
a \qty{1}{\uM} Jacalin test solution 
reveals a concentration of \qty{1.4e5}{particles\per\milli\liter} (Fig.~\ref{fig:scatter}(a)).
The observed
distribution of size and refractive index
values is consistent with expectations for protein aggregates \cite{wang2016protein}.
Results for a \qty{0.5}{\uM} solution of human IgA are
reported in Fig.~\ref{fig:scatter}(b), 
and similarly indicate a small concentration of dispersed particles,
\qty{2.8e5}{particles\per\milli\liter},
which we also ascribe to 
protein aggregates. 
The agglutination assay is initiated by incubating a \qty{1}{\uM} Jacalin test solution
with a \qty{0.2}{\uM} IgA sample. The result of this assay in Fig.~\ref{fig:scatter}(c) mimics the high end of the average
physiological range (at 100\texttimes\ dilution)
and yields a signal of \qty{1.4e7}{particles\per\milli\liter}, which we ascribe to protein agglutinates.
This concentration of
detectable particles was over ten-folds greater than background
while remaining
within the operational range
of xSight.

\subsection{Particle Concentration}

The observed particle concentration depends strongly on which 
protein is present in the sample with Jacalin.
The data in Fig.~\ref{fig:particleconcentration}
were obtained by incubating samples 
at specified concentrations
with \qty{0.5}{\uM} Jacalin.
Human IgA incubation with Jacalin showed a positive correlation between 
the concentration of added IgA and the number of particles counted by THC measurements.
Variability in measured particle concentration
is roughly three times larger for samples containing
IgA agglutinates.
The number density of detected particles
in the IgA samples
increases linearly with the 
initial concentration
of dissolved IgA over the range
of concentrations from \qty{5}{\mgdl} to
\qty{30}{\mgdl}.
This coincides with the clinically
relevant reference interval, taking into account 
the ten-fold dilution in the protocol.
No such trend is apparent in the 
trials with human IgG or 
in the negative control measurements with ADH. 
For both IgG and ADH, the observed range of particle concentrations
is consistent with instrumental uncertainty.

\begin{figure*}[ht]
\centering
\includegraphics[width=\textwidth]{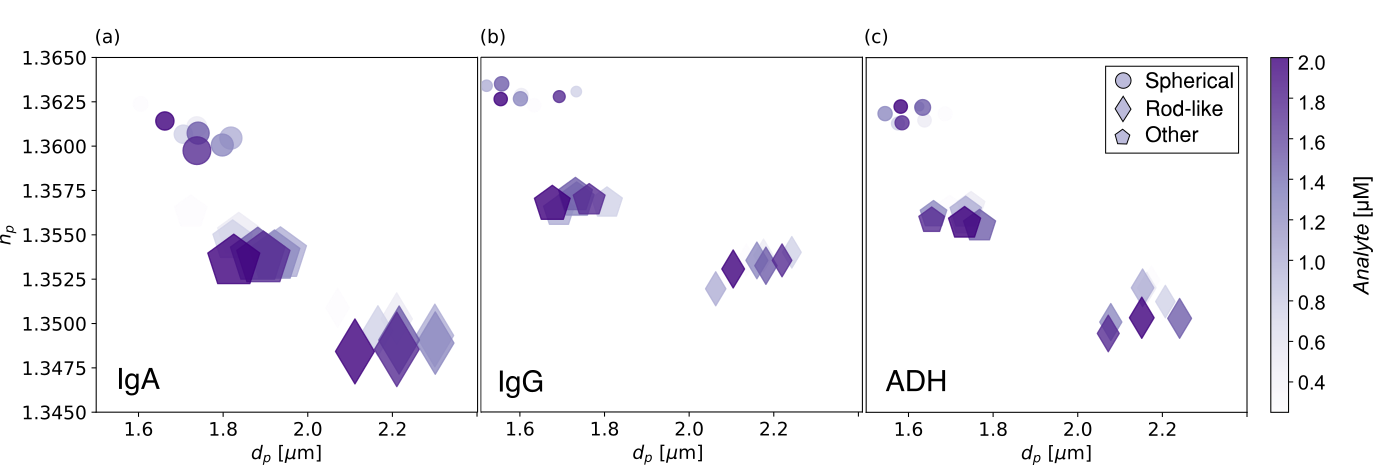}
\caption[]{Diameter and refractive index of Jacalin agglutinates by particle morphology. 
Spherical, rod-like and other-shaped agglutinates formed after incubation of Jacalin with human serum IgA (a) and IgG (b). 
Negative control with alcohol dehydrogenase from yeast (c). 
All experiments with immunoglobulins were conducted in triplicates. 
The size of the marker represents the average particle count.} 
	\label{fig:morph_IgA_IgG_ADH}
\end{figure*}

\subsection{Particle Morphology}
The symmetry of the hologram was defined by a qualitative classification of the morphology of the observed agglutinate 
and sorted into three categories: Spherical, Rod-like and Other.
Spherical particles are identified automatically by their radially symmetric interference patterns.
Similarly, Rod-like particles create holograms with biaxial symmetry.
More ramified particles create asymmetric holograms and accordingly are placed in a separate category. 
The smallest distribution of morphology for all analytes was for spherical agglutinates. The majority of particles that contribute to the particle concentration in the agglutination were found to be either rod-like or other morphology, as seen in Fig.~\ref{fig:morph_IgA_IgG_ADH}. When the holograms of the agglutinates are distinguished by their morphology, the rod-like and other shaped particles show the greatest difference in the diameter, refractive index and particle concentration for samples that contain IgA compared to those with IgG. Particles with a spherical morphology in all samples are generally unaffected by the amount of antibody present.

\subsection{Particle Size}

The apparent diameter ($d_p$) is obtained for each particle imaged in the sample by the digital camera.
All particles larger than \qty{500}{\nm} in the observation volume are detected and analyzed by the
instrument.
Smaller particles are detected with lower efficiency.
The diameter provides an indicator of the apparent size of each protein agglutinate. 
As IgA concentration was increased, the average diameters of Jacalin-IgA agglutinates 
did not vary proportionally with increasing IgA concentration, as seen from Fig.~\ref{fig:morph_IgA_IgG_ADH}. 

\subsection{Particle Refractive Index}
Lorenz-Mie analysis yields values for the refractive index of each particle, $n_p$,
that can be used to infer additional information about agglutinates, including
the concentration of proteins within each particle.
This inference is performed using effective-medium theory \cite{markel2016introduction} starting from the refractive index of the medium, $n_m$, and
the intrinsic refractive index of the protein, $n_0$.
We compute a protein's refractive index from the sequence of amino acids
obtained from the Protein Databank (PDB) database \cite{vonHofe2025}.
A protein sequence composed of $N$-amino acids yields a set of $N$ refractive index values,
$\{n_j\}$ at the imaging wavelength \cite{mcmeekin1964refractive,zhao_distribution_2011}.
The Lorentz-Lorenz factor for that sequence is
\begin{equation}
    L = \sum_{j = 1}^N \phi_j \, \frac{n_j^2 - 1}{n_j^2 + 2},
\end{equation}
where $\phi_j = v_j / \sum_{i = 1}^N v_i$ is the volume fraction
of the protein occupied by amino acid $j$ given
its specific volume, $v_j$ \cite{mcmeekin1964refractive}.
The predicted refractive index for the protein is then
\begin{equation}
    n_0 = \frac{2 L + 1}{1 - L} .
\end{equation}
By the same line of reasoning, the volume fraction of protein in the
effective sphere enclosing a protein agglomerate is obtained from the
agglomerate's measured refractive index as
\begin{equation}
    \phi = \frac{n_p^2 - n_m^2}{n_p^2 + 2 n_m^2} \frac{n_0^2 + 2 n_m^2}{n_0^2 - n_m^2} .
\end{equation}

The results are included in the supporting information Table~\ref{tbl:index_increments}. There was a distinct correlation between the refractive index of the particle and its morphology, as seen in 
Fig.~\ref{fig:morph_IgA_IgG_ADH}.
The spherical particles had the highest refractive index, at \num{1.363},
indicating the presence of a greater concentration of protein. The rod-like particles had the lowest apparent refractive index at \num{1.347}.
Since rod-like and irregular particles are analyzed with the effective-sphere model \cite{altman2021holographic}
the reported refractive index includes a contribution from the aqueous medium and therefore
is lower than the bulk refractive index of the protein itself.
The downward trend in refractive index in Fig.~\ref{fig:morph_IgA_IgG_ADH} from spherical to rod-like
agglutinates reflects the increasing proportion of buffer accommodated by the increasingly
open morphology.

\section{Discussion}

The lectin, Jacalin, is selective for galactose containing glycoproteins \cite{arockia_jeyaprakash_structural_2003}, 
allowing us to utilize this lectin to demonstrate the first proof of concept of a Holographic Agglutination Assay (HAA). 
Jacalin contains four binding sites for galactose and is known to bind to IgA. 
This multi-valent binding interaction can result in protein agglutination composed of fractal assemblies with an anticipated ratio of 4:1.
We selected the range of concentration of IgA to model clinical assays 
that monitor IgA levels in Human serum as indicators of pathological conditions\cite{puissant-lubrano_immunoglobulin_2015,gonzalez-quintela_serum_2008}.
Experiments in this study are representative of a lectin assay that can be performed over the entire clinical reference interval
in a single microfluidic xCell with a single lectin. 
The amount of agglutinates produced in a 10x diluted emulated patient sample report on IgA levels in the sample,
as low (\qty{0.5}{\uM}) or high (\qty{2}{\uM}).

\subsection{Quantifying Agglutination using Total Holographic Characterisation}

The agglutination assays conducted for this study relied on the simplicity of the sample composition to confirm that the agglutinates formed and observed by holographic characterisation are due to an agglutinated particle composed of the lectin, Jacalin, and the corresponding antibody. 
The visible particles were assumed to be a result of fractal assemblies of the Jacalin selectively binding to the IgA. 
These assembled agglutinates are visible because (i) the refractive index of the proteins in the agglutinates is higher compared to the buffer medium, ($n_m = \num{1.34}$) as seen in Table~\ref{tbl:index_increments} 
and (ii) the agglutinates can grow over the incubation time to become large enough to scatter the wavelength of light used in the study, \qty{450}{\nm}. 
Despite the similarity in the refractive index of all proteins used in this assay,
it was possible to distinguish between samples that contained IgA or IgG due to the agglutinate concentration in each measurement. 
We hypothesised that the only additional particles formed after incubation with the analyte protein 
that could reasonably be present in the sample would be a result of protein agglutination. 
Therefore, by comparing two types of Human Immunoglobulins, 
we could quantify the agglutinates in the sample by correlating the presence of visible agglutinates
to selectivity of the immunoglobulin towards Jacalin. 
Upon observation of an increase in particle concentration after addition of the target analyte,
we assume that the particle count represents the number of Jacalin-IgA agglutinates in the sample. 
The range of concentrations of IgA and IgG were kept consistent to allow a quantitative comparison of the immunoglobulins.

The background level of \qty{e6}{particles\per\milli\liter} 
can be ascribed to protein aggregates in the stock solution. 
Since the assay was performed using serum antibodies, there may also be a trace amount of serum proteins present.
To ensure that the assay is quantitative, the small number of protein aggregates in the stock solutions 
can be subtracted from the total particle concentration of the analyte.
This accounts for any protein aggregates formed as a result of non-specific binding 
or the expected denaturation of protein molecules in the solution over time. 
THC can thus clearly detect protein agglutination 
as a result of selective binding above background.

\subsection{Agglutination as a measure of IgA concentration}

The IgA and IgG antibody isotypes had a different response to the presence of Jacalin in the sample. 
The agglutinates were found to be proportionally abundant for higher concentrations of IgA, 
Fig.~\ref{fig:particleconcentration}. The increased range of particle concentrations
is consistent with run-to-run variability in the agglutination process \cite{dolgosheina1992kinetic}.
Compared to IgA, Human IgG incubation with Jacalin did not show any increase in Jacalin-IgG agglutinates with increasing IgG concentration. This result demonstrates the capacity of the agglutination assay to quantify selectivity of the analyte protein towards Jacalin.
Alcohol dehydrogenase from yeast (ADH1) was used as a negative control because of its comparable molecular weight to an immunoglobulin, \qty{150}{\kilo\dalton}, and its lack of glycosylation sites.
Similar to the effect of incubation with IgG, increasing the concentration of ADH in the assay did not increase the number of Jacalin agglutinates measured.
Measurements with IgA and Jacalin showed a clear positive correlation between the increasing IgA concentration and the particle count. This supports the presence of an increasing number of IgA antibodies with the terminal galactose bound to the four galactose-specific carbohydrate binding sites of Jacalin leading to a higher distribution of agglutinates. 
In contrast, the apparently low levels of agglutinates in measurements of samples with IgG or ADH support the conclusion that Jacalin preferentially agglutinates with IgA, not with IgG or ADH. 
The negative result with ADH confirms the
absence of non-specific agglutination by
Jacalin.

\subsection{Morphology of lectin-Ig agglutinates}

Holographic characterisation allows analysis of particle morphology. Each particle is categorised into three categories: Spherical, Rod-like or Other.
This provides an additional parameter for interpreting the properties derived from the hologram of each particle, such as $d_p$ and $n_p$, and the sample as a whole.
In this study,
morphological classification provides a unique perspective into agglutinates of lectins and glycoproteins.
Comparing results of Jacalin assays with human serum IgA against IgG as shown in Fig.~\ref{fig:morph_IgA_IgG_ADH}, it is evident that when we consider all three variables, $d_p$, $n_p$, and the total particle count, the difference in selectivity of the Jacalin towards IgA is easy to identify by the distributions independent of morphology. 
Unlike in previously reported holographic assays, the diameter of the agglutinates was not proportional to the analyte concentration. This observation was not surprising, as the diameter of the agglutinate is not expected to depend on the concentration of available analyte. 
The size of the plot symbols in Fig.~\ref{fig:morph_IgA_IgG_ADH} is proportional to the number
of observations in each morphological category and therefore to the number density of the
associate agglutinates.
Comparing symbol sizes in Fig.~\ref{fig:morph_IgA_IgG_ADH} therefore highlights the selectivity
of the assay.
Not only are substantially more agglutinates formed in samples with IgA, but a greater proportion
of those agglutinates adopt rod-like or other complex morphologies that are consistent with expectations
for multipoint agglutination by a multivalent lectin.
This suggests that particles with non-spherical morphology may be preferentially monitored to determine whether the glycoprotein in the sample is selective for the lectin. 
By using all morphology types currently classified by holographic characterisation, we can obtain a more detailed view of the different forms of protein agglutinates in a sample. 
This provides a significant advantage over other assays that would allow our proof-of-concept agglutination assay to better describe the correlation of the analyte concentration and the level of measured agglutinates. 
By selectively monitoring the diameter, refractive index and particle count of a particular morphology of protein agglutinate, it is possible to get a characteristic fingerprint for the correlation between the lectin and a biochemically relevant glycoprotein.

\section{Conclusions}

We have demonstrated the first example of a lectin agglutination assay using holographic characterisation.
This approach can quantify the agglutination of human serum antibodies with Jacalin, a lectin that binds specifically to terminal galactose. 
The reported work takes advantage of the label-free holographic characterisation technique to design a quantitative holographic assay that does not require a probe bead. 
The IgA-specific lectin Jacalin used in this study was observed to only cause agglutination in samples with IgA and had no such effect when only the lectin or the analyte was present.
The results from the assay also demonstrate the selectivity of the Jacalin for IgA, one of the two most abundant isotypes of antibodies in human serum, over IgG.
Taken together, the positive correlation of particle concentration with IgA and the negative result with IgG 
illustrate the selectivity of the lectin-based assay for IgA versus another immunoglobulin. 
We demonstrate that the number of protein agglutinates formed in the presence of the lectin could be used to quantify the amount of IgA present. 
The holographic agglutination assays highlight the difference between the proteins that can be induced to agglutinate by Jacalin, like IgA, 
compared to those that may simply share a similar size and refractive index, like ADH. 
ADH was used as a simple negative control protein for immunoglobulins because of its similar molecular weight and lack of glycosylation sites. 
Particle morphology of protein agglutinates could be determined from the holograms. 
This provided an additional qualitative classification of the formed agglutinates that informs about the complexity resulting from variations in size, refractive index and particle concentration in a sample. 

The reported Holographic Agglutination Assay can be used to quantify the levels of IgA in an emulated biological sample.
The results of this work indicate that this assay could be extended to monitoring various immunoglobulins using lectin agglutination, without the inclusion of a probe bead. 
Although only a single lectin must be present in the assays reported to ensure that the result is a useful metric of agglutination, 
we can unambiguously demonstrate the selectivity of the lectin.
Future work could improve the analytical capacity of this assay by exploring other known ligands of Jacalin, such as IgD.
This assay may also be used with highly specific lectins that recognise only a limited set of glycosylation patterns similar to lectin arrays.  
We envision the development of this assay to monitor clinical samples where the level of agglutination from various lectins at regular time intervals would provide clinically useful information about the presence of a diseased state, its progression over time or the effect of medical treatments on the patient's immunoglobulin levels.

\section*{Author Contributions}
R.Q.: Conceptualisation, design and validation of methodology, experimental investigation and visualisation. 
K.K.: Supervision, conceptualisation, methodology validation and interpretation of data, and funding acquisition. 
D.G.G.: Supervision, conceptualisation, formal analysis and interpretation of data and funding acquisition. 
All authors contributed to the writing, editing and reviewing the manuscript.

\section*{Conflicts of interest}
D.G.G.\ is a founder of Spheryx, Inc., which manufactures xSight, including
the instrument used for this study.

\section*{Acknowledgements}
The initial stage of this work was supported by the RAPID
program of the National Science Foundation through
Award No.~DMR-2027013.
Additional support was provided by the NSF through Awards No.~DMR-2104837 
and Award No.~DMR-2438983.
The Spheryx xSight used for this study was acquired
as shared instrumentation by the NYU MRSEC under support
of the NSF through Award No.~DMR-1420073.
Dan Wasserburg helped with preliminary measurements of the agglutination experiments.

\section{Appendix: Protein refractive indexes}

\begin{table}[h]
	\caption{Effective refractive index of proteins used in the Holographic Agglutination Assay. Calculations were done using protein refractive index increments, $dn/dc$, with the Lorentz-Lorenz factor for the sequence of amino acids obtained from the Protein Databank.}
	\label{tbl:index_increments}
    \centering
	\begin{tabular}{l l c c}
		\hline
		Protein & PDB & $dn/dc~(\si{\milli\liter\per\gram})$ & $n_\text{protein}$ \\
		\hline
        Jacalin $\alpha$ chain & 1KU8 & 0.1820 & 1.628  \\
		Jacalin $\beta$ chain & 1KU8 & 0.1890 & 1.640  \\
		Human IgG & 1HZH & 0.1849 & 1.701  \\
		Human IgA & 1IGA & 0.1831 & 1.699 \\
		\hline
	\end{tabular}
\end{table}

\bibliography{agglutination} 

\begin{thebibliography}{10}
\newcommand{\enquote}[1]{``#1''}

\bibitem{lee2007characterizing}
S.-H. Lee, Y.~Roichman, G.-R. Yi, \emph{et~al.}, \enquote{Characterizing and tracking single colloidal particles with video holographic microscopy,} {\protect\JournalTitle{Opt. Express}} \textbf{15}, 18275--18282 (2007).

\bibitem{martin2022inline}
C.~Martin, L.~E. Altman, S.~Rawat, \emph{et~al.}, \enquote{In-line holographic microscopy with model-based analysis,} {\protect\JournalTitle{Nat. Rev. Methods Primers}} \textbf{2}, 83 (2022).

\bibitem{krishnatreya2014measuring}
B.~J. Krishnatreya, A.~Colen-Landy, P.~Hasebe, \emph{et~al.}, \enquote{Measuring {Boltzmann}'s constant through holographic video microscopy of a single colloidal sphere,} {\protect\JournalTitle{Am. J. Phys.}} \textbf{82}, 23--31 (2014).

\bibitem{cheong2011holographic}
F.~C. Cheong, K.~Xiao, D.~J. Pine, and D.~G. Grier, \enquote{Holographic characterization of individual colloidal spheres' porosities,} {\protect\JournalTitle{Soft Matter}} \textbf{7}, 6816--6819 (2011).

\bibitem{odete2020role}
M.~A. Odete, F.~C. Cheong, A.~Winters, \emph{et~al.}, \enquote{The role of the medium in the effective-sphere interpretation of holographic particle characterization data,} {\protect\JournalTitle{Soft Matter}} \textbf{16}, 891--898 (2020).

\bibitem{wang2016holographic}
C.~Wang, F.~C. Cheong, D.~B. Ruffner, \emph{et~al.}, \enquote{Holographic characterization of colloidal fractal aggregates,} {\protect\JournalTitle{Soft Matter}} \textbf{12}, 8774--8780 (2016).

\bibitem{altman2020interpreting}
L.~E. Altman and D.~G. Grier, \enquote{Interpreting holographic molecular binding assays with effective medium theory,} {\protect\JournalTitle{Biomed. Opt. Express}} \textbf{11}, 5225--5236 (2020).

\bibitem{fung2013holographic}
J.~Fung and V.~N. Manoharan, \enquote{Holographic measurements of anisotropic three-dimensional diffusion of colloidal clusters,} {\protect\JournalTitle{Phys. Rev. E}} \textbf{88}, 020302 (2013).

\bibitem{wang2014using}
A.~Wang, T.~G. Dimiduk, J.~Fung, \emph{et~al.}, \enquote{Using the discrete dipole approximation and holographic microscopy to measure rotational dynamics of non-spherical colloidal particles,} {\protect\JournalTitle{J. Quant. Spectr. Rad. Trans.}} \textbf{146}, 499--509 (2014).

\bibitem{wang2016protein}
C.~Wang, X.~Zhong, D.~B. Ruffner, \emph{et~al.}, \enquote{Holographic characterization of protein aggregates,} {\protect\JournalTitle{J. Pharm. Sci.}} \textbf{105}, 1074--1085 (2016).

\bibitem{altman2021holographic}
L.~E. Altman, R.~Quddus, F.~C. Cheong, and D.~G. Grier, \enquote{Holographic characterization and tracking of colloidal dimers in the effective-sphere approximation,} {\protect\JournalTitle{Soft Matter}} \textbf{17}, 2695--2703 (2021).

\bibitem{altman2023anomalous}
L.~E. Altman, A.~D. Hollingsworth, and D.~G. Grier, \enquote{Anomalous tumbling of colloidal ellipsoids in {Poiseuille} flows,} {\protect\JournalTitle{Phys. Rev. E}} \textbf{108}, 034609 (2023).

\bibitem{winters2020quantitative}
A.~Winters, F.~C. Cheong, M.~A. Odete, \emph{et~al.}, \enquote{Quantitative differentiation of protein aggregates from other subvisible particles in viscous mixtures through holographic characterization,} {\protect\JournalTitle{J. Pharm. Sci.}} \textbf{109}, 2405--2412 (2020).

\bibitem{abdulali2022multi}
R.~Abdulali, L.~E. Altman, and D.~G. Grier, \enquote{Multi-angle holographic characterization of individual fractal aggregates,} {\protect\JournalTitle{Opt. Express}} \textbf{30}, 38587--38595 (2022).

\bibitem{zagzag2020holographic}
Y.~Zagzag, M.~F. Soddu, A.~D. Hollingsworth, and D.~G. Grier, \enquote{Holographic molecular binding assays,} {\protect\JournalTitle{Sci. Rep.}} \textbf{10}, 1932 (2020).

\bibitem{snyder_quddus_2020}
K.~Snyder, R.~Quddus, A.~D. Hollingsworth, \emph{et~al.}, \enquote{Holographic immunoassays: direct detection of antibodies binding to colloidal spheres,} {\protect\JournalTitle{Soft Matter}} \textbf{16}, 10180--10186 (2020).

\bibitem{cheong2009flow}
F.~C. Cheong, B.~S.~R. Dreyfus, J.~Amato-Grill, \emph{et~al.}, \enquote{Flow visualization and flow cytometry with holographic video microscopy,} {\protect\JournalTitle{Opt. Express}} \textbf{17}, 13071--13079 (2009).

\bibitem{kasimbeg_holographic_2019}
P.~N.~O. Kasimbeg, F.~C. Cheong, D.~B. Ruffner, \emph{et~al.}, \enquote{Holographic characterization of protein aggregates in the presence of silicone oil and surfactants,} {\protect\JournalTitle{J. Pharm. Sci.}} \textbf{108}, 155--161 (2019).

\bibitem{middleton_optimizing_2019}
C.~Middleton, M.~D. Hannel, A.~D. Hollingsworth, \emph{et~al.}, \enquote{Optimizing the synthesis of monodisperse colloidal spheres using holographic particle characterization,} {\protect\JournalTitle{Langmuir}} \textbf{35}, 6602--6609 (2019).

\bibitem{rahn2023strengths}
H.~Rahn, M.~Oeztuerk, N.~Hentze, \emph{et~al.}, \enquote{The strengths of total holographic video microscopy in detecting sub-visible protein particles in biopharmaceuticals: A comparison to {Flow Imaging} and {Resonant Mass Measurement},} {\protect\JournalTitle{J. Pharm. Sci.}} \textbf{112}, 985--990 (2023).

\bibitem{bedrossian_quantifying_2017}
M.~Bedrossian, C.~Barr, C.~A. Lindensmith, \emph{et~al.}, \enquote{Quantifying microorganisms at low concentrations using digital holographic microscopy ({DHM}),} {\protect\JournalTitle{{JoVE}}} p. e56343 (2017-11-01).

\bibitem{luo_quantitative_2021}
Y.~Luo, H.-A. Joung, S.~Esparza, \emph{et~al.}, \enquote{Quantitative particle agglutination assay for point-of-care testing using mobile holographic imaging and deep learning,} {\protect\JournalTitle{Lab Chip}} \textbf{21}, 3550--3558 (2021).

\bibitem{polpanich2007detection}
D.~Polpanich, P.~Tangboriboonrat, A.~Elaissari, and R.~Udomsangpetch, \enquote{Detection of malaria infection via latex agglutination assay,} {\protect\JournalTitle{Anal. Chem.}} \textbf{79}, 4690--4695 (2007).

\bibitem{stavitsky_agglutination_1998}
A.~B. Stavitsky, \enquote{Agglutination,} in \emph{Encyclopedia of Immunology (Second Edition),}  P.~J. Delves, ed. (Elsevier, 1998-01-01), pp. 56--59.

\bibitem{arrhenius_agglutination_1908}
S.~Arrhenius, \enquote{On agglutination and coagulation.} {\protect\JournalTitle{Journal of the American Chemical Society}} \textbf{30}, 1382--1388 (1908).

\bibitem{malomgre_recent_2009}
W.~Malomgré and B.~Neumeister, \enquote{Recent and future trends in blood group typing,} {\protect\JournalTitle{Anal. Bioanal. Chem.}} \textbf{393}, 1443--1451 (2009).

\bibitem{li_blood_2022}
H.-Y. Li and K.~Guo, \enquote{Blood group testing,} {\protect\JournalTitle{Front. Med.}} \textbf{9}, 827619 (2022).

\bibitem{widal1896serodiagnostic}
F.~Widal, \enquote{Serodiagnostic de la fi\`{e}vre typho\"{\i}de,} {\protect\JournalTitle{Bull. M\'{e}m. Soc. M\'{e}d. H\^{o}p. Paris}} \textbf{13}, 561--566 (1896).

\bibitem{olopoenia_widal_2000}
L.~A. Olopoenia and A.~L. King, \enquote{Widal agglutination test -- 100 years later: still plagued by controversy,} {\protect\JournalTitle{Postgrad. Med. J.}} \textbf{76}, 80--84 (2000).

\bibitem{hirst1942quantitative}
G.~K. Hirst, \enquote{The quantitative determination of influenza virus and antibodies by means of red cell agglutination,} {\protect\JournalTitle{J. Exp. Med.}} \textbf{75}, 49 (1942).

\bibitem{esmail_rapid_2021}
S.~Esmail, M.~J. Knauer, H.~Abdoh, \emph{et~al.}, \enquote{Rapid and accurate agglutination-based testing for {SARS}-{CoV}-2 antibodies,} {\protect\JournalTitle{Cell Rep. Methods}} \textbf{1}, 100011 (2021).

\bibitem{ribeiro_dot_2013}
J.~P. Ribeiro and L.~K. Mahal, \enquote{Dot by dot: analyzing the glycome using lectin microarrays,} {\protect\JournalTitle{Curr. Opin. Chem. Biol.}} \textbf{17}, 827--831 (2013).

\bibitem{rademacher1988glycobiology}
T.~Rademacher, R.~Parekh, and R.~Dwek, \enquote{Glycobiology,} {\protect\JournalTitle{Annu. Rev. Biochem.}} \textbf{57}, 785--838 (1988).

\bibitem{kabir_Jacalin_1998}
S.~Kabir, \enquote{Jacalin: a jackfruit (\emph{Artocarpus heterophyllus}) seed-derived lectin of versatile applications in immunobiological research,} {\protect\JournalTitle{J. Immunol. Methods}} \textbf{212}, 193--211 (1998).

\bibitem{arockia_jeyaprakash_structural_2003}
A.~Arockia~Jeyaprakash, S.~Katiyar, C.~P. Swaminathan, \emph{et~al.}, \enquote{Structural basis of the carbohydrate specificities of {Jacalin}: An x-ray and modeling study,} {\protect\JournalTitle{J. Mol. Biol.}} \textbf{332}, 217--228 (2003).

\bibitem{roque-barreira_Jacalin_1985}
M.~C. Roque-Barreira and A.~Campos-Neto, \enquote{Jacalin: an {IgA}-binding lectin.} {\protect\JournalTitle{J. Immunol.}} \textbf{134}, 1740--1743 (1985). Number: 3.

\bibitem{backer_autoantibodies_1989}
E.~T. Backer and G.~A. Harff, \enquote{Autoantibodies to lactate dehydrogenase in serum identified by use of immobilized protein {G} and immobilized jacalin, a jackfruit lectin,} {\protect\JournalTitle{Clin. Chem.}} \textbf{35}, 2190--5 (1989).

\bibitem{pack_purification_2000}
T.~D. Pack, \enquote{Purification of human {IgA},} {\protect\JournalTitle{Curr. Protoc. Immunol.}} \textbf{38}, 2.10B.1--2.10B.7 (2000).

\bibitem{andre_impairment_1990}
P.~M. Andre, P.~Le~Pogamp, and D.~Chevet, \enquote{Impairment of jacalin binding to serum {IgA} in {IgA} nephropathy,} {\protect\JournalTitle{J. Clin. Lab. Anal.}} \textbf{4}, 115--119 (1990).

\bibitem{ma_serum_2020}
H.~Ma, W.~Zeng, H.~He, \emph{et~al.}, \enquote{Serum {IgA}, {IgM}, and {IgG} responses in {COVID}-19,} {\protect\JournalTitle{Cell. Mol. Immunol.}} \textbf{17}, 773--775 (2020).

\bibitem{allez_immunodeficiency_2004}
M.~Allez and L.~Mayer, \enquote{Immunodeficiency,} in \emph{Encyclopedia of Gastroenterology,}  L.~R. Johnson, ed. (Elsevier, 2004-01-01), pp. 425--429.

\bibitem{baenziger_structure_1974}
J.~Baenziger and S.~Kornfeld, \enquote{Structure of the carbohydrate units of {IgA}1 immunoglobulin: {II},} {\protect\JournalTitle{J. Biol. Chem.}} \textbf{249}, 7270--7281 (1974).

\bibitem{skea_studies_1988}
D.~L. Skea, P.~Christopoulous, A.~G. Plaut, and B.~J. Underdown, \enquote{Studies on the specificity of the {IgA}-binding lectin, jacalin,} {\protect\JournalTitle{Mol. Immunol.}} \textbf{25}, 1--6 (1988).

\bibitem{delacroix_iga_1982}
D.~L. Delacroix, C.~Dive, J.~C. Rambaud, and J.~P. Vaerman, \enquote{{IgA} subclasses in various secretions and in serum.} {\protect\JournalTitle{Immunology}} \textbf{47}, 383--385 (1982).

\bibitem{parkkinen1989aberrant}
J.~Parkkinen, \enquote{Aberrant lectin-binding activity of immunoglobulin {G} in serum from rheumatoid arthritis patients.} {\protect\JournalTitle{Clin. Chem.}} \textbf{35}, 1638--1643 (1989).

\bibitem{alter_antibody_2018}
G.~Alter, T.~H. Ottenhoff, and S.~A. Joosten, \enquote{Antibody glycosylation in inflammation, disease and vaccination,} {\protect\JournalTitle{Semin. Immunol.}} \textbf{39}, 102--110 (2018).

\bibitem{yoann_2016}
F.~S. van~de Bovenkamp, L.~Hafkenscheid, T.~Rispens, and Y.~Rombouts, \enquote{The emerging importance of {IgG} {Fab} glycosylation in immunity,} {\protect\JournalTitle{J. Immunol.}} \textbf{196}, 1435--1441 (2016).

\bibitem{aucouturier_characterization_1987}
P.~Aucouturier, E.~Mlhaesco, C.~Mihaesco, and J.-L. Preud'homme, \enquote{Characterization of jacalin, the human {IgA} and {IgD} binding lectin from jackfruit,} {\protect\JournalTitle{Mol. Immunol.}} \textbf{24}, 503--511 (1987).

\bibitem{whyte2018normal}
M.~B. Whyte and P.~Kelly, \enquote{The normal range: it is not normal and it is not a range,} {\protect\JournalTitle{Postgrad. Med. J.}} \textbf{94}, 613--616 (2018).

\bibitem{dieguez2020serum}
M.~Dieguez-Alvarez, I.~Carballo, M.~Alonso-Sampedro, \emph{et~al.}, \enquote{Serum immunoglobulin-{A} ({IgA}) concentrations in a general adult population: association with demographics and prevalence of selective {IgA} deficiency,} {\protect\JournalTitle{Clin. Chem. Lab. Med.}} \textbf{58}, e109--e112 (2020).

\bibitem{markel2016introduction}
V.~A. Markel, \enquote{Introduction to the {Maxwell Garnett} approximation: tutorial,} {\protect\JournalTitle{J. Opt. Soc. Am. A}} \textbf{33}, 1244--1256 (2016).

\bibitem{vonHofe2025}
J.~von Hofe, J.~Abacousnac, M.~Chen, \emph{et~al.}, \enquote{Multivalency controls the growth and dynamics of a biomolecular condensate,} {\protect\JournalTitle{J. Am. Chem. Soc.}} \textbf{147}, 25242–25253 (2025).

\bibitem{mcmeekin1964refractive}
T.~L. McMeekin, M.~L. Groves, and N.~J. Hipp, \enquote{Refractive indices of amino acids, proteins, and related substances,} in \emph{Amino Acids and Serum Proteins,}  J.~Stekol, ed. (ACS Publications, Washington, DC, 1964), chap.~4, pp. 54--66.

\bibitem{zhao_distribution_2011}
H.~Zhao, P.~H. Brown, and P.~Schuck, \enquote{On the distribution of protein refractive index increments,} {\protect\JournalTitle{Biophys. J.}} \textbf{100}, 2309--2317 (2011).

\bibitem{puissant-lubrano_immunoglobulin_2015}
B.~Puissant-Lubrano, M.~Peres, P.-A. Apoil, \emph{et~al.}, \enquote{Immunoglobulin {IgA}, {IgD}, {IgG}, {IgM} and {IgG} subclass reference values in adults,} {\protect\JournalTitle{Clin. Chem. Lab. Med.}} \textbf{53}, e359--e361 (2015-11-01).

\bibitem{gonzalez-quintela_serum_2008}
A.~Gonzalez-Quintela, R.~Alende, F.~Gude, \emph{et~al.}, \enquote{Serum levels of immunoglobulins ({IgG}, {IgA}, {IgM}) in a general adult population and their relationship with alcohol consumption, smoking and common metabolic abnormalities,} {\protect\JournalTitle{Clin. Exp. Immunol.}} \textbf{151}, 42--50 (2008).

\bibitem{dolgosheina1992kinetic}
E.~B. Dolgosheina, A.~Y. Karulin, and A.~V. Bobylev, \enquote{A kinetic model of the agglutination process,} {\protect\JournalTitle{Math. Biosci.}} \textbf{109}, 1--10 (1992).

\end{thebibliography}

\end{document}